\documentclass[3p,times,procedia]{elsarticle}

\usepackage{ecrc}

\usepackage{ulem}

\volume{00}

\firstpage{1}

\runauth{}

\usepackage{amssymb}

\usepackage[figuresright]{rotating}

\begin{document}

\begin{frontmatter}

\dochead{}
\title{Testing the Pauli Exclusion Principle for electrons at LNGS}

\author[label1]{H.~Shi\corref{cor1}}
\ead{Hexi.Shi@oeaw.ac.at}
\author[label2]{S.~Bartalucci}
\author[label3]{S.~Bertolucci}
\author[label1,label2]{C.~Berucci}
\author[label2,label4]{A.M.~Bragadireanu}
\author[label1]{M.~Cargnelli}
\author[label2]{A.~Clozza}
\author[label2,label4,label5]{C.~Curceanu}
\author[label2]{L.~De Paolis}
\author[label6]{S.~Di Matteo}
\author[label2]{A.~d'Uffizi}
\author[label7]{J.-P.~Egger}
\author[label2]{C.~Guaraldo}
\author[label2]{M.~Iliescu}
\author[label1]{T.~Ishiwatari}
\author[label1]{J.~Marton\corref{cor2}}
\author[label8]{M.~Laubenstein}
\author[label9]{E.~Milotti}
\author[label2,label4]{D.~Pietreanu}
\author[label2,label5]{K.~Piscicchia}
\author[label6]{T.~Ponta}
\author[label10]{A.~Romero Vidal}
\author[label2]{E.~Sbardella}
\author[label2]{A.~Scordo}
\author[label2,label4]{D.L.~Sirghi}
\author[label2,label4]{F.~Sirghi}
\author[label2]{L.~Sperandio}
\author[label11]{O.~Vazquez Doce}
\author[label1]{E.~Widmann}
\author[label1]{J.~Zmeskal}

\address[label1]{Stefan-Meyer-Institut f\"{u}r Subatomare Physik, Boltzmanngasse 3, 1090 Wien, Austria,}
\address[label2]{INFN, Laboratori Nazionali di Frascati, C.P. 13, Via E. Fermi 40, I-00044 Frascati(Roma), Italy,}
\address[label3]{CERN, CH-1211, Geneva 23, Switzerland,}
\address[label4]{IFIN-HH, Institutul National pentru Fizica si Inginerie Nucleara Horia Hulubbei, Reactorului 30, Magurele, Romania,}
\address[label5]{Museo Storico della Fisica e Centro Studi e Ricerche Enrico Fermi, Piazza del Viminale 1, 00183 Roma, Italy,}
\address[label6]{Institut de Physique UMR CNRS-UR1 6251, Universit\'{e} de Rennes1, F-35042 Rennes, France,}
\address[label7]{Institut de Physique, Universit\'{e} de Neuch\^{a}tel, 1 rue A.-L. Breguet, CH-2000 Neuch\^{a}tel, Switzerland,}
\address[label8]{INFN, Laboratori Nazionali del Gran Sasso, S.S. 17/bis, I-67010 Assergi (AQ), Italy,}
\address[label9]{Dipartimento di Fisica, Universit\'{a} di Trieste and INFN-Sezione di Trieste, Via Valerio, 2, I-34127 Trieste, Italy,}
\address[label10]{Universidade de Santiago de Compostela, Casas Reais 8, 15782 Santiago de Compostela, Spain,}
\address[label11]{Excellence Cluster Universe, Technische Universit\"{a}t M\"{u}nchen, Boltzmannstra\ss e 2, D-85748 Garching, Germany.}

\cortext[cor2]{Presenter}
\cortext[cor1]{Corresponding author}


\begin{abstract}
 High-precision experiments have been done to test the Pauli exclusion principle (PEP) for electrons 
 by searching for anomalous $K$-series X-rays from a Cu target supplied with electric current. 
 With the highest sensitivity, the VIP (VIolation of Pauli Exclusion Principle) experiment 
 set an upper limit at the level of $10^{-29}$ for the probability that an external electron captured by a Cu atom 
 can make the transition from the 2$p$ state to a 1$s$ state already occupied by two electrons. 
 In a follow-up experiment at Gran Sasso, we aim to increase the sensitivity by two orders of magnitude.
 We show proofs that the proposed improvement factor is realistic based on the results from recent performance 
 tests of the detectors we did at Laboratori Nazionali di Frascati (LNF). 
 }
\end{abstract}
 

\end{frontmatter}

\graphicspath{{Figures/}}

\section{Introduction}
\label{intro}
Formulated by Wolfgang Pauli in 1925 \cite{Pau25}, 
the Pauli Exclusion Principle (PEP) is one of the building blocks of quantum mechanics. 
It is the foundation for our understanding of nuclear physics, particle physics, condensed-matter physics, and astrophysics 
where many-fermion systems are concerned.
By far the principle does not have an intuitive explanation for its physical cause, 
and there is possibility that high precision experiment may discover small violation that could reveal more fundamental principles.
However the experimental test is difficult because there has been no well-established parameter that can account for PEP violation quantitatively in a theory, 
and that can also be derived from experiments for direct comparison.

In the reviews given by Greenberg and Mohapatra \cite{GreM89,Gre89}, 
they surveyed over the searches for a phenomenology framework for possible small violation of the PEP, 
and then pointed out that no satisfactory solution could be found to be consistent 
within a local 
field theory. 
However, they argued that following the parameterization proposed by Ignatiev and Kuzmin (IK) \cite{Ign87}, 
in 
an extended 
model of a single fermion-like oscillator which allows double occupancy with a small amplitude of $\beta$, 
one can discuss about the phenomenology of a small violation of the PEP with a parameter quantitatively derivable from experiments. 

The first precision measurement done by Ramberg and Snow \cite{Ram90} follows a method 
Greenberg and Mohapatra \cite{Gre87} proposed after they extended the IK model. 
The method first used by Goldhaber and Scharff-Goldhaer \cite{Gol48} back in 1948, 
was original intended to check if the beta ray from beta decay is identical to ordinary electrons. 
Their ingenius idea was that, if not identical to electrons, 
the beta rays absorbed by a block of metal (in this case lead) will neglect all the electrons occupying the atomic states and deexcite via the cascade process. 
The 2$p$-1$s$ transition that violates the PEP will have different energy from the normal $2p$ - $1s$ transition due to 
the shielding effect of an additional electron in the ground state \cite{Cur13}. 
Based on the non-existence of the anomalous X-rays, they first concluded the equivalence of beta ray to electron, 
and more interestingly they later pointed out the experiment can be interpreted as a test for PEP. 
A quantitative evaluation based on the result of the experiment was done by Greenberg \cite{Gre89}, 
who deduced explicitly that the possibility that the PEP can be violated is less than 0.03.

The idea of introducing external ``fresh" electrons to the target system as applied by this pioneering experiment and the Ramberg-Snow experiment 
is crucial in the method of testing the PEP for electrons.
Without the ``fresh" electrons, 
two experiments in the 1970s \cite{Rei74, Log79} looked for prohibited X-rays or $\gamma$ rays from stable atomic or nucleus systems, 
and argued the null results served as tests for the PEP violation.
However this type of measurement does not validate to be a test 
because it has assumed that the transitions between different permutation group could occur, 
Such assumption violates in the first place the 
superselection rule separating states in different presentations of the symmetric group \cite{Mes64}.
On the other hand, 
external electrons that had no interaction with the target system not only make source of electrons in large population possible, 
they are also the prerequisite that 
small violation of the PEP can be discussed in the framework of quantum mechanics as Greenberg proposed \cite{Gre89}. 
Becasue the newly captured electron and the copper atom have the possiblility of forming a ``mixed" symmetry state that is highly excited, 
from which anomalous X-rays can be observed. 
Without introducing external electrons, 
anomalous X-rays can not be observed since the possible ``mixed" symmetry state will always be at its ground state.
To represent the probability of a small violation of the PEP in the absence of a field theory, 
Ramberg and Snow used the $\beta$ parameter introduced first in the IK model. 
For a random pair of electrons, 
$1 - \frac12\beta^2$ is the possibility of the pair in the normal antisymmetric state, 
and $\frac12\beta^2$ the probability in the anomalous symmetric state.
In the IK model, $\beta$ is explicitly defined with the zero, one, and two particle states of $|0\rangle$, $|1\rangle$, and $|2\rangle$, 
together with the creation operator $a^{\dagger}$ and the annihilation operator $a$ as : 
\begin{eqnarray}
a^{\dagger} |0\rangle = |1\rangle, a^{\dagger} |1\rangle = \beta\ |2\rangle, a^{\dagger} |2\rangle = 0; \ \ \ \nonumber 
a\ |0\rangle = 0, a\ |1\rangle = |0\rangle, a\ |2\rangle = \beta\ |1\rangle.\nonumber
\end{eqnarray}
Following the IK model, Greenberg and collaborators constructed the ``quon" algebra \cite{Gre91} with $q$ parameter: 
\begin{eqnarray}
a_{k} a_{l}^{\dagger} - q\ a_{l}^{\dagger} a_{k} = \delta_{kl}, \nonumber
\end{eqnarray}
which can be understood as the average of the Bose and Fermi commutation relations:
\begin{eqnarray}
\frac{1+q}{2}[a_{k}, a_{l}^{\dagger}]_{-} + \frac{1-q}{2}[a_{k}, a_{l}^{\dagger}]_{+} = \delta_{kl}, \nonumber
\end{eqnarray}
and the $\beta$ parameter can be written in terms of the $q$ parameter as:
\begin{eqnarray}
\frac12\ \beta^2 = \frac12\ (1+q). \nonumber
\end{eqnarray}
Although still having open questions to solve, 
the ``quon" theory is by far the best attempt to violate by a small amount the Fermi and the Bose statistics. 
However for a direct comparison of the experimental results, 
the VIP experiment used and will use the same notation of $\beta^2$ by Ramberg and Snow.

In next section, 
we describe the details of the experimental method used by VIP experiment,
and show the improvements in the sensitivity achieved in past experiments. 
Afterwards we introduce our follow-up experiment of VIP2 and its progress in the ongoing preparation.

\section{ Experimental method } 

With the same idea of searching for anomalous transition X-rays, 
Ramberg and Snow improved drastically the sensitivity 
by changing the source of electrons from beta decay to constant electric current. 
Performed at the ground floor of the Muon building at Fermilab, 
they used a proportional tube counter as the X-ray detector with a resolution of 1 keV at 8 keV, 
and a large array of plastic scintillators to veto possible signals from charged cosmic rays.
A thin strip copper as target was connected to a 50 A power supply. 
By comparing the X-ray spectra from measurements with and without power supply, 
the excess of events in the forbiden transition energy region when current is supplied, 
will in principle be violation to the PEP. 

\begin{figure}[htb]
\centering
\includegraphics[width=12cm,clip]{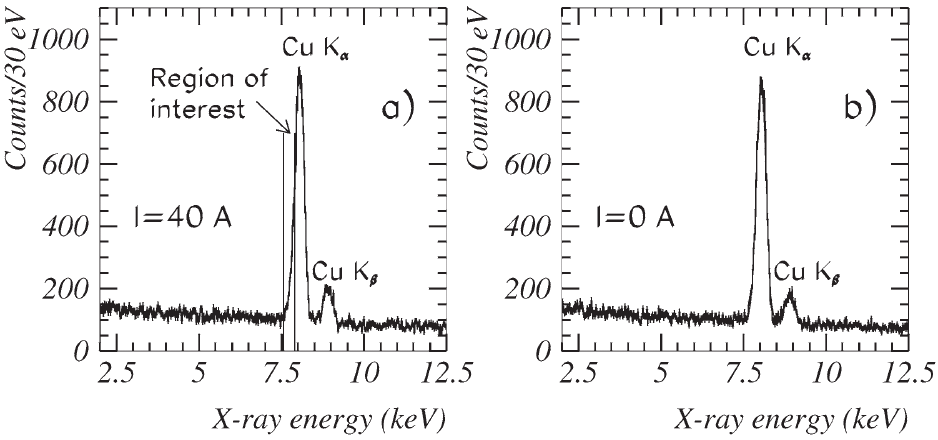}
\caption{
         Energy spectra for the VIP experiment \cite{Bar06} : (a) with 40 A current, (b) without current, from part of the data set.
         Normal K transitions of Cu present are background, 
         and they are due to excitation of target by cosmic rays and environmental radiations. 
         }
\label{fig-0}       
\end{figure}

\subsection{VIP experiment and results}
\begin{table}[h!]
\caption{Limits of the Pauli violation probability for electrons from recent high precision experiments: 
        }
\label{tab:pastexp}
\begin{center}
\begin{tabular}[b]{l l l l}
\hline
\hline
Experiment &  Target &   Upper limit of $\beta^2$/2  & reference \\
\hline
Ramberg-Snow        & Copper &  1.7 $\times$ 10 $^{-26}$  & \cite{Ram90} \\
S.R. Elliott et al. & Lead   &  1.5 $\times$ 10 $^{-27}$  & \cite{Ell12} \\
VIP(2006)           & Copper &  4.5 $\times$ 10 $^{-28}$  & \cite{Bar06} \\
VIP(2012)           & Copper &  4.7 $\times$ 10 $^{-29}$  & \cite{Cur12} \\
VIP2(goal)          & Copper &      $\times$ 10 $^{-31}$  & \cite{Mar13} \\
\hline\hline
\end{tabular}
\end{center}
\end{table}

VIP experiment followed the similar method of Ramberg-Snow experiment, 
and used the same definition of the parameter $\frac12\beta^2$ to represent the violation to the PEP for a direct comparison of the experiment results. 
The improvement in sensitivity was achieved 
firstly due to the site of the experiment at the underground laboratory in Laboratori Nazionali del Gran Sasso (LNGS), 
which has the advantage of the excellent shielding against cosmic rays \cite{Bar06}. 
The other reason is the use of Charge Coupled Device (CCD) as the X-ray detector which had a typical resolution of 320 eV at 8 keV, 
that increased the precision in the definition of the region of interest to search for anomalous X-rays. 

In Table \ref{tab:pastexp}, all the results from experiments using ``fresh" electrons are listed, 
together with the goal of the planned VIP2 experiment at LNGS.

\subsection{VIP2 experiment}
\subsubsection{Design}

\begin{figure}[htb]
\centering
\includegraphics[width=12cm,clip]{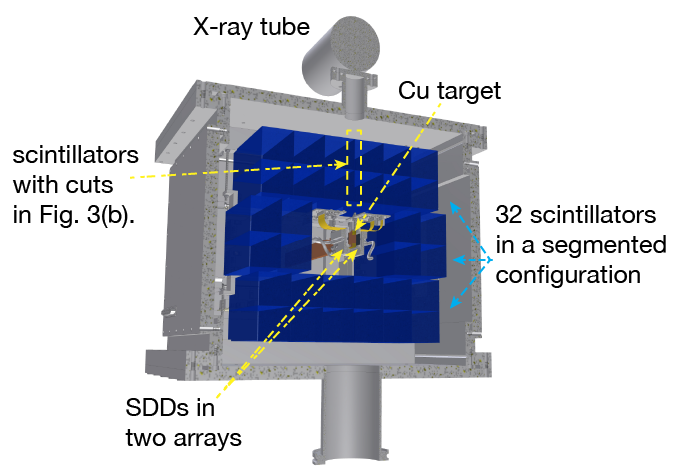}
\caption{
         An artist presentation for the cutaway view of the setup. 
         Over 90\% of the solid angle for the SDDs acceptance is covered by 32 plastic scintillators as active shielding.
         The timing capability of the SDDs will allow us to reduce most of the background of the Cu $K$-series X-rays 
         induced by cosmic rays impinging on the target. 
         }
\label{fig-1}       
\end{figure}

In the follow-up experiment of VIP2, 
we aim to improve the sensitivity of VIP experiment by two orders of magnitude \cite{Mar11}. 
A detailed list for the features that will contribute to the overall improvement is summarized in Table \ref{tab:improvement} \cite{Mar13}. 
The dominant factor of background reduction will come from the application of the Silicon Drift Detectors (SDDs) as the X-ray detector 
and their active shielding using arrays of plastic scintillator as veto counters. 
Comparing to the readout time at the order of seconds for CCD,  
the SDD has a charge collection time of less than one micro-second. 
This allows us to use the time correlation between the X-ray events and the events at the veto counters, 
to exclude all the X-rays, including the $K$-series X-rays of Cu from the target excited by cosmic rays or by the environmental radiation, 
as the energy spectra in Fig. \ref{fig-0} show.

We plan to use six SDD detectors with a total active area of 6 cm$^2$ mounted close to the pure Cu target in the shape of a strip 3 cm in length. 
Surrounding the SDDs and readout electronics, as shown in Fig. \ref{fig-1}, 
32 pieces of plastic scintillators each with a dimension of 40 mm $\times$ 32 mm $\times$ 250 mm will be mounted in a segmented configuration, 
covering about 90\% of the solid angle for the acceptance of SDDs. 
To readout the light output of each scintillator, 
we attach with optic cement solid-state Silicon Photo-Multipliers (SiPMs) directly to the scintillators. 
More information for the plastic scintillator and SiPM and its electronic board can be found in Ref. \cite{Mar13}.

\begin{table}[htbp]
\caption{
         The improvement factors for VIP2 in comparison to the features of VIP \cite{Mar13} : 
        }
\label{tab:improvement}
\begin{center}
\begin{tabular}[b]{l l l}
\hline
\hline
Changes in VIP2     &  value VIP2 (VIP)  &    expected gain  \\
\hline
acceptance          & 12 \%           &  12  \\
increase current    & 100 A (40 A)    & $>$ 2   \\
reduced length      & 3 cm (8.8 cm )  &  1/3 \\
\hline
total linear factor &              &     8  \\
\hline 
energy resolution &  170 eV (320 eV) @ 8 keV   &  4 \\   
reducecd active area  & 6 cm $^2$ (114 cm $^2$   &   20 \\
better shielding and veto   &         &    5-10 \\
higher SDD efficiency   &   &    1/2 \\
\hline 
background reduction  &  &  200 - 400 \\
\hline 
overall improvement  &  &   $>$ 120 \\
\hline
\end{tabular}
\end{center}
\end{table}

Since the precise information about the energy deposit by charged particles or environmental radiations in the scintillators is not our interest, 
we plan to take the time over threshold (ToT) information of the SiPM's signal instead of the QDC data. 
Such an application is of technical interest, 
about which we show later some promising results from the test measurement of cosmic rays.

\subsubsection{ Beam test of scintillators as veto counters }

\begin{figure}[h!]
\centering
\includegraphics[width=12cm,clip]{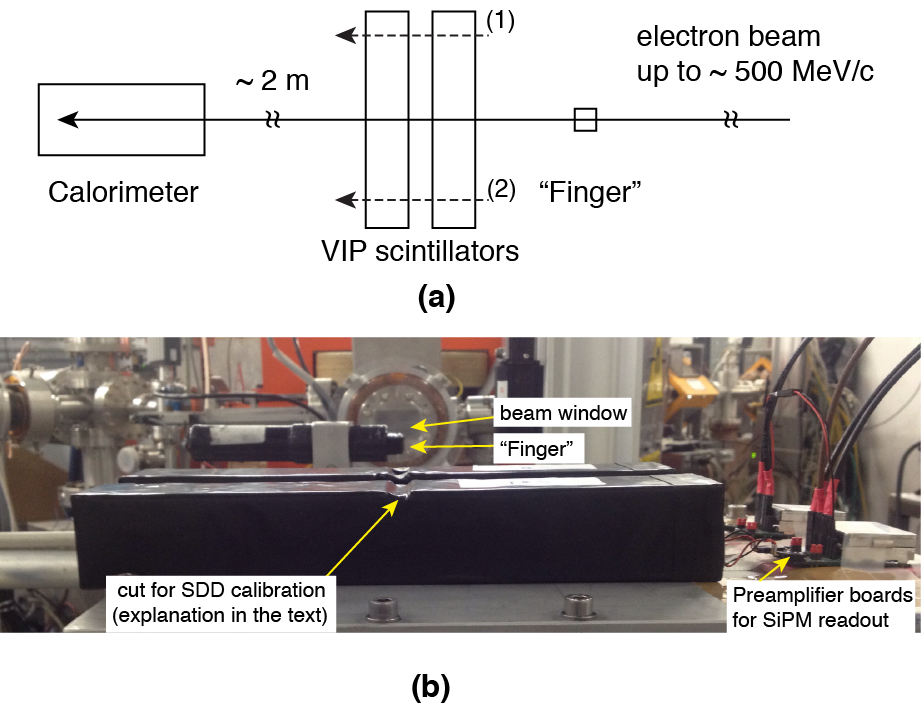}
\caption{
         Figures show the BTF test for the scintillators with cuts to allow the X-ray tube to irradiate the pure metal foils 
         close to the SDDs for energy calibration. 
         To confirm the unique structure does not affect the performance of the scintillators, 
         the efficiency and timing resolution were tested with electron beam in the LNF beam test factory (BTF) in December 2013.
         Figure (a) shows the schematic layout of the beam test configuration. 
         A small scintillator of one cubic cm in size ("Finger") at the upper stream and a calorimeter at down stream of the beam 
         are used to define the timing and to choose the single electron events.
         Two different hit positions of the beam are denoted by (1) and (2). 
         Figure (b) is a photo taken from the beam direction, down stream of the VIP2 scintillators, 
         showing the configuration of beam hitting the center position of the scintillators. 
        }
\label{fig-2}       
\end{figure}

For the energy calibration of SDDs during the data taking, 
we plan to use a X-ray tube mounted to one window of the setup to radiate thin foils of pure metals placed near the SDDs.
To let the radiation reach the foils, 
four scintillators between the SDDs and the tube are prepared with special cuts, which can be seen from Fig. \ref{fig-2} (b). 

To test the performance of the scintillators together with the readout system, 
we did a test measurement using 500 MeV/c electron beam at the Beam Test Facility (BTF) at LNF in December 2013. 
The experiment layout and a photo taken from the downstream of the beam are shown in Fig. \ref{fig-2}. 
We took data for three different hit positions of the beam along the long direction of the scintillators. 
The results show that, 
the dependence of light yield on the hit position of the charged particle does not affect the efficiency, 
which is constantly above 97\%. 
We confirmed the time resolution of the scintillator is about 2 to 3 ns as shown in \ref{fig-4} (a). 
This resolution is small enough for our application to use timing of scintillators 
to mark the start timing for the SDD events with a spread in time of hundreds of nanoseconds.

\subsubsection{ Test setup in LNF }

\begin{figure}[htb]
\centering
\includegraphics[width=12cm,clip]{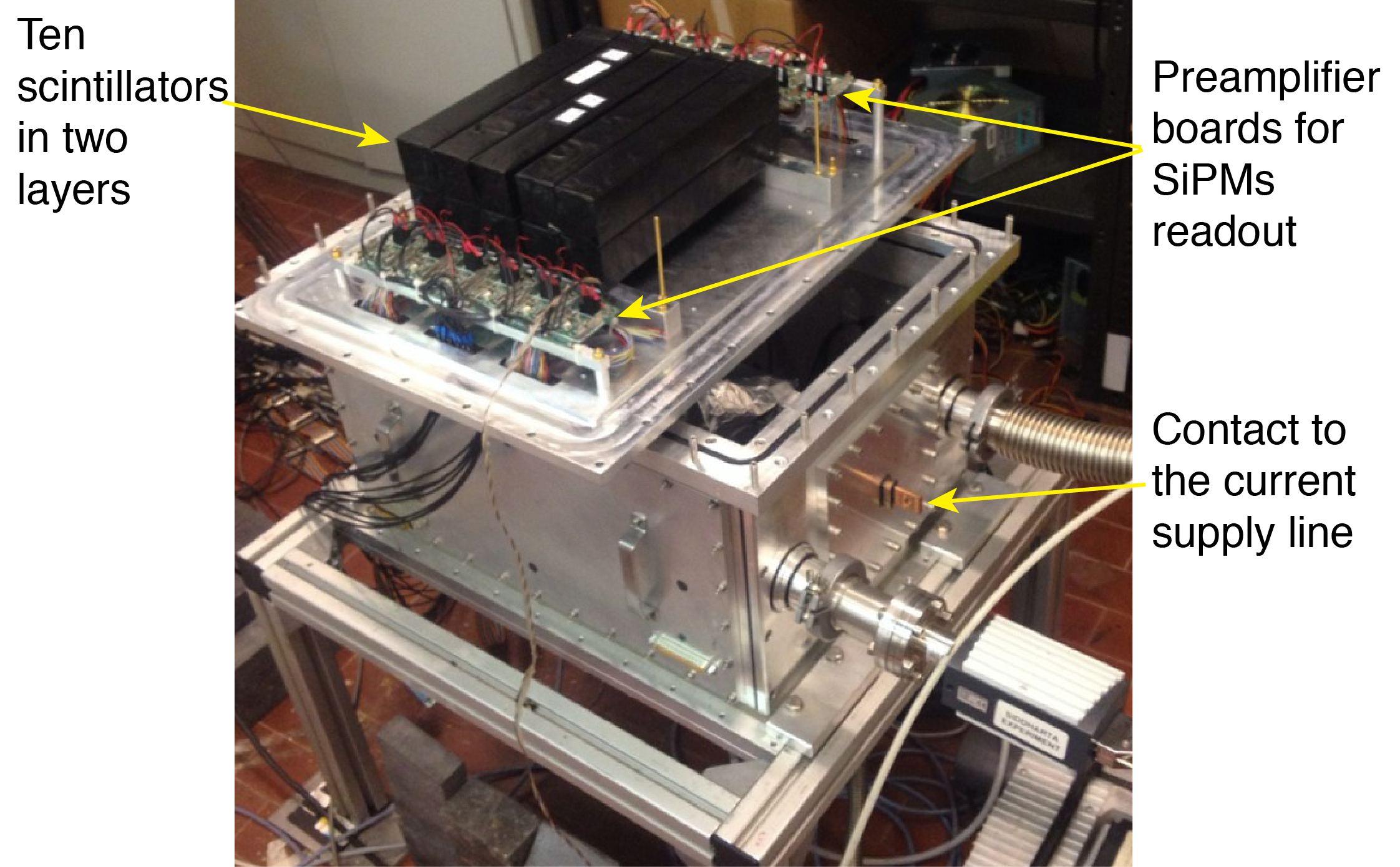}
\caption{
         A photo of the test setup before it is closed for cooling test. 
         Ten scintillators in two layers with their readout system are mounted on the lid. 
         }
\label{fig-3}       
\end{figure}

Before mounting the whole setup to LNGS for the long-term measurement, 
we are testing its performance and stability of the vacuum and the cryogenics system with a setup in the laboratory at LNF-INFN. 
The six SDDs in two arrays are mounted in the setup, 
and ten of the total 32 scintillators to be used as veto counters are mounted in two layers above the SDDs. 
Fig. \ref{fig-3} is a photo of the test setup before we closed the lid cover, 
to which the top-layer scintillators are mounted. 

To reach the working condition required by the SDDs, 
we use a helium compressor to cool the setup with a set temperature of about 80 Kelvin. 
Then the SDDs are further cooled to about - 140 $\,^{\circ}\mathrm{C}$ by an aluminum pipe line in which Ar gas is circulated.  
The pressure of the setup is kept at the level of 10$^{-5}$ mbar with a turbo pump.

\begin{figure}[htb]
\centering
\includegraphics[width=16cm,clip]{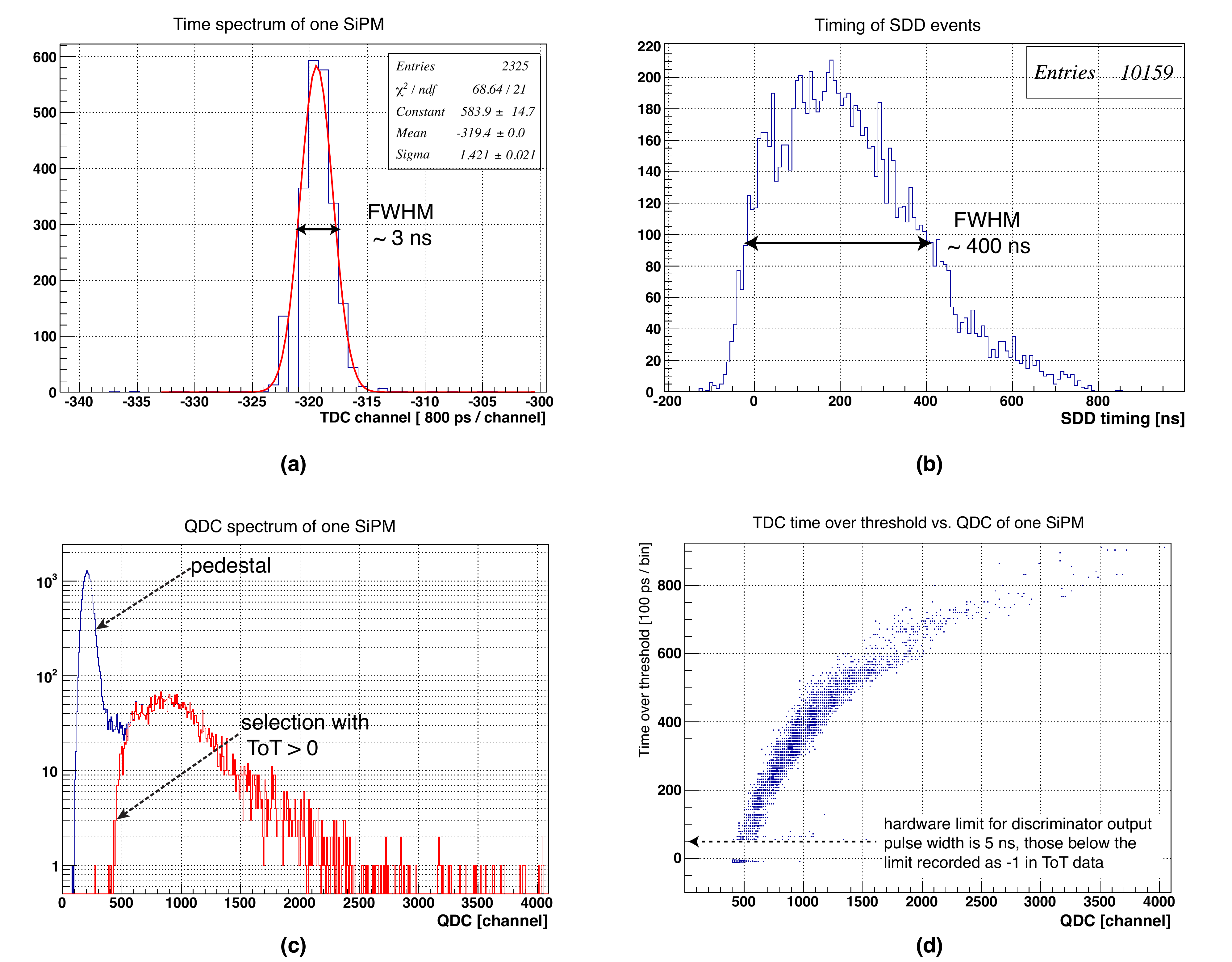}
\caption{
         Preliminary results from the beam test in BTF and measurements of cosmic ray with test setup. 
         (a) shows the time spectrum of one SiPM on the scintillator from the BTF test, resolution is about 3 ns FWHM;
         (b) shows the time spectrum of five working SDDs at the temperature of -130 $\,^{\circ}\mathrm{C}$ to - 150 $\,^{\circ}\mathrm{C}$, with a FWHM of about 400 ns;
         (c) is a typical QDC spectrum from one SiPM. The red histogram here shows the selection with time over threshold cut, 
         which excluded the pedestal events clearly; 
         (d) for the same SiPM in (c), this figure shows the correlation between the QDC and the time over threshold. 
         }
\label{fig-4}       
\end{figure}

With this test setup we measured cosmic rays to check the trigger scheme as well as the timing performance of the SDDs. 
A cosmic ray event and its timing are defined by the coincidence of hit events at both two layers of scintillators.  
The timing of the X-ray events at the SDDs that are correlated to the cosmic ray events are shown in Fig. \ref{fig-4} (b). 
At the temperature of -130 $\,^{\circ}\mathrm{C}$ to - 150 $\,^{\circ}\mathrm{C}$,    
the time spectrum from the summation of five SDDs shows a FWHM of about 400 ns, 
and the shape of the distribution in timing has the feature shown by the same SDDs used in SIDDHARTA experiment \cite{Baz11}. 
Comparing to the time resolution for the SDDs, 
the uncertainty in the start timing indicated by the time resolution of on SiPM shown in Fig. \ref{fig-4} (a), 
is negligibly small.

Finally, we confirmed from this test the feasibility of using the ToT information for the SiPM readout instead of QDC data. 
In the test setup, we took both QDC data and the ToT information from TDC. 
From the QDC data, 
we see that selecting the events with none-zero ToT value, 
the cosmic ray hit events can be separated from the pedestal as shown in Fig. \ref{fig-4} (c). 
Moreover, the time over threshold of the SiPM preamplifier signal shows clear correlation to the QDC of the signal in Fig. \ref{fig-4} (d). 
The QDC distribution over 2000 channels corresponds to 1000 channels of ToT, 
indicating the possible application of ToT instead of QDC. 
This will give us more freedom in the analysis for the Gran Sasso measurement data in which only TDC information will be recorded.

As a summary for our test measurements in BTF and in the laboratory, 
we conclude that the efficiency of the scintillators and the time rosolution of the SDDs are confirmed to be capable to 
achieve the estimatedestimated  background reduction level in the proposal of VIP2 experiment.

\section{Summary and outlook}
Recent high-precision experiments make it possible to discuss the phenomenological implications of small violations of PEP. 
The VIP experiment at the Gran Sasso underground laboratory performed X-ray spectroscopy 
to search for Pauli-principle forbiden transitions of electrons from a Cu strip,   
when it is supplied with external electrons by means of electric current. 
For this type of experiment, 
we achieved the highest sensitivity in setting the limit for possible small violation of PEP which can occur for electrons. 

In our follow-up experiment of VIP2 which is also planned in Gran Sasso, 
we aim to improve the sensitivity by two orders of magnitude in setting the limit for the violation of PEP. 
The performances of the setup and detectors are under test at LNF at the point of early 2014, 
and the results confirm the feasibility of the expected improvement factor of background reduction.

\paragraph{Acknowledgements}
We thank H. Schneider, L. Stohwasser, and D. St\"{u}ckler from Stefan-Meyer-Institut 
for their fundamental contribution in designing and building the VIP2 setup. 
We thank the very important assistance of the INFN-LNGS laboratory staff during all phases of preparation, 
installation and data taking as well as the support from the HadronPhysics FP6(506078), 
HadronPhysics2 FP7 (227431), HadronPhysics3 (283286) projects and the EU COST 1006 Action is gratefully acknowledged. 
Especially we thank the Austrian Science Foundation (FWF) which supports the VIP2 project with the grant P25529-N20, 
and we thank the support from the EU COST Action MP1006, Fundamental Problems in Quantum Physics, 
and from Centro Fermi (''Problemi aperti nella meccania quantistica" project).


\bibliographystyle{elsarticle-num}

\end{document}